\PassOptionsToPackage{unicode}{hyperref}
\PassOptionsToPackage{hyphens}{url}
\PassOptionsToPackage{dvipsnames,svgnames,x11names}{xcolor}
\documentclass[
  12pt]{article}

\usepackage{amsmath,amssymb}
\usepackage{iftex}
\ifPDFTeX
  \usepackage[T1]{fontenc}
  \usepackage[utf8]{inputenc}
  \usepackage{textcomp} % provide euro and other symbols
\else % if luatex or xetex
  \usepackage{unicode-math}
  \defaultfontfeatures{Scale=MatchLowercase}
  \defaultfontfeatures[\rmfamily]{Ligatures=TeX,Scale=1}
\fi
\usepackage{lmodern}
\ifPDFTeX\else  
\fi
\IfFileExists{upquote.sty}{\usepackage{upquote}}{}
\IfFileExists{microtype.sty}{% use microtype if available
  \usepackage[]{microtype}
  \UseMicrotypeSet[protrusion]{basicmath} % disable protrusion for tt fonts
}{}
\makeatletter
\@ifundefined{KOMAClassName}{% if non-KOMA class
  \IfFileExists{parskip.sty}{%
    \usepackage{parskip}
  }{% else
    \setlength{\parindent}{0pt}
    \setlength{\parskip}{6pt plus 2pt minus 1pt}}
}{% if KOMA class
  \KOMAoptions{parskip=half}}
\makeatother
\usepackage{xcolor}
\makeatletter
\ifx\paragraph\undefined\else
  \let\oldparagraph\paragraph
  \renewcommand{\paragraph}{
    \@ifstar
      \xxxParagraphStar
      \xxxParagraphNoStar
  }
  \newcommand{\xxxParagraphStar}[1]{\oldparagraph*{#1}\mbox{}}
  \newcommand{\xxxParagraphNoStar}[1]{\oldparagraph{#1}\mbox{}}
\fi
\ifx\subparagraph\undefined\else
  \let\oldsubparagraph\subparagraph
  \renewcommand{\subparagraph}{
    \@ifstar
      \xxxSubParagraphStar
      \xxxSubParagraphNoStar
  }
  \newcommand{\xxxSubParagraphStar}[1]{\oldsubparagraph*{#1}\mbox{}}
  \newcommand{\xxxSubParagraphNoStar}[1]{\oldsubparagraph{#1}\mbox{}}
\fi
\makeatother

\providecommand{\tightlist}{%
  \setlength{\itemsep}{0pt}\setlength{\parskip}{0pt}}\usepackage{longtable,booktabs,array}
\usepackage{calc} % for calculating minipage widths
\usepackage{etoolbox}
\makeatletter
\patchcmd\longtable{\par}{\if@noskipsec\mbox{}\fi\par}{}{}
\makeatother
\IfFileExists{footnotehyper.sty}{\usepackage{footnotehyper}}{\usepackage{footnote}}
\makesavenoteenv{longtable}
\usepackage{graphicx}
\makeatletter
\def\maxwidth{\ifdim\Gin@nat@width>\linewidth\linewidth\else\Gin@nat@width\fi}
\def\maxheight{\ifdim\Gin@nat@height>\textheight\textheight\else\Gin@nat@height\fi}
\makeatother
\setkeys{Gin}{width=\maxwidth,height=\maxheight,keepaspectratio}
\makeatletter
\def\fps@figure{htbp}
\makeatother

\makeatletter
\@ifpackageloaded{caption}{}{\usepackage{caption}}
\AtBeginDocument{%
\ifdefined\contentsname
  \renewcommand*\contentsname{Table of contents}
\else
  \newcommand\contentsname{Table of contents}
\fi
\ifdefined\listfigurename
  \renewcommand*\listfigurename{List of Figures}
\else
  \newcommand\listfigurename{List of Figures}
\fi
\ifdefined\listtablename
  \renewcommand*\listtablename{List of Tables}
\else
  \newcommand\listtablename{List of Tables}
\fi
\ifdefined\figurename
  \renewcommand*\figurename{Figure}
\else
  \newcommand\figurename{Figure}
\fi
\ifdefined\tablename
  \renewcommand*\tablename{Table}
\else
  \newcommand\tablename{Table}
\fi
}
\@ifpackageloaded{float}{}{\usepackage{float}}
\floatstyle{ruled}
\@ifundefined{c@chapter}{\newfloat{codelisting}{h}{lop}}{\newfloat{codelisting}{h}{lop}[chapter]}
\floatname{codelisting}{Listing}

\makeatother
\makeatletter
\@ifpackageloaded{caption}{}{\usepackage{caption}}
\@ifpackageloaded{subcaption}{}{\usepackage{subcaption}}
\makeatother

\ifLuaTeX
  \usepackage{selnolig}  % disable illegal ligatures
\fi
\usepackage[]{natbib}
\usepackage{bookmark}

\IfFileExists{xurl.sty}{\usepackage{xurl}}{} % add URL line breaks if available
\hypersetup{
  pdftitle={Title},
  pdfauthor={Author 1; Author 2},
  pdfkeywords={3 to 6 keywords, that do not appear in the title},
  colorlinks=true,
  linkcolor={blue},
  filecolor={Maroon},
  citecolor={Blue},
  urlcolor={Blue},
  pdfcreator={LaTeX via pandoc}}

\usepackage{booktabs}
\usepackage{tabularx}
\usepackage{hyperref}
\usepackage[utf8]{inputenc}
\usepackage{graphicx}
\usepackage{color}
\usepackage{lineno}
\usepackage{multirow}
\usepackage{rotating}
\usepackage{arydshln}
\usepackage{amsmath} % For the equation and math symbols
\usepackage{amsfonts} % For math symbols if needed
\usepackage{algorithm} % For the floating environment
\usepackage{algpseudocode} % For the pseudocode commands
\newcommand{\anon}{1}

\begin{document}

\def\spacingset#1{\renewcommand{\baselinestretch}%
{#1}\small\normalsize} \spacingset{1}

%%%%%%%%%%%%%%%%%%%%%%%%%%%%%%%%%%%%%%%%%%%%%%%%%%%%%%%%%%%%%%%%%%%%%%%%%%%%%%

\if1\anon
{
  \title{\bf A spatial random forest algorithm for population-level epidemiological risk assessment}
  \author{Duncan Lee\thanks{
    Corresponding author email - Duncan.Lee@glasgow.ac.uk}\hspace{.2cm}\\
    School of Mathematics and Statistics, University of Glasgow\\
    and \\
    Vinny Davies \\
    School of Mathematics and Statistics, University of Glasgow}
  \maketitle
} \fi

\if0\anon
{
  \bigskip
  \bigskip
  \bigskip
  \begin{center}
    {\LARGE\bf Title}
\end{center}
  \medskip
} \fi

\bigskip
\begin{abstract}
Spatial epidemiology identifies the drivers of elevated population-level disease risks, using disease counts, exposures and known confounders at the areal unit level. Poisson regression models are typically used for inference, which incorporate a linear/additive regression component and allow for unmeasured confounding via a set of spatially autocorrelated random effects. This approach requires the confounder interactions and their functional relationships with disease risk to be specified in advance, rather than being learned from the data. Therefore, this paper proposes the SPAR-Forest-ERF algorithm, which is the first fusion of random forests for capturing non-linear and interacting confounder-response effects with Bayesian spatial autocorrelation models that can estimate interpretable exposure response functions (ERF) with full uncertainty quantification. Methodologically, we extend existing methods set in a prediction context by propagating uncertainty between both the ML and statistical models, developing a new stopping criteria designed  to ensure the stability of  the primary inferential target, and incorporating a range of different ERFs for maximum model flexibility. This methodology is motivated by a new study quantifying the impact of air pollution concentrations on self-rated health in Scotland, using data from the recently released 2022 national census.
\end{abstract}

\noindent%
\vfill

\newpage
% \spacingset{1} % DON'T change the spacing! - changed for arxiv

\section{Introduction}\label{sec-intro}
An important area of spatial epidemiology is quantifying the impacts that exposures have on the risk of disease, with examples including the impact of air pollution on Covid-19 hospitalization rates (\citealp{lee2022}), and  the association between the accessibility of green space and mental wellbeing (\citealp{nutsford2013}). In these studies the region of interest is partitioned into $K$ non-overlapping small areal units, and the data for each unit summarize the total number of disease cases, average exposures, and relevant confounding factors for the population who live there. While these studies cannot estimate individual-level cause and effect due to their population-level nature, their advantages lie in their geographical coverage of a large population at risk, and the ease in obtaining the required non-disclosive data that makes them fast and inexpensive to implement. 

Bayesian hierarchical models are often used to make inference from these data, whose mean function depends on a population offset to control for demography, the exposures of interest, a set of confounders and  spatially autocorrelated random effects. The latter capture any remaining spatial autocorrelation in the data not explained by the covariates, due to issues such as unmeasured confounding. Conditional autoregressive (CAR, \citealp{besag1991}) prior distributions are commonly assigned to these random effects, while linear or simple additive relationships are assumed between the confounders and disease risk. This modeling framework provides interpretable exposure-response functions (ERF) with full uncertainty quantification (UQ), and has the ability to capture residual spatial autocorrelation. However, confounder interactions and their functional relationships with disease risk have to be specified in advance by the analyst, which requires a large number of assumptions and prevents these patterns being learned from the data. 

Machine learning (ML) algorithms such as random forests and neural networks capture complex and interacting confounder-response relationships, but they assume independence between the observations and are hence unsuitable for spatial data. They also provide limited inferential outputs such as variable importance plots, which are much less epidemiologically relevant than relative risk based ERFs that arise naturally from statistical models. A number of authors have extended  ML models for spatial data, and recent reviews are given by \cite{wikle2023}  and \cite{Patelli2024}. The simplest approach represents the spatial autocorrelation in the data with spatially smooth quantities such as basis functions (\citealp{chen2024}), or spatially lagged covariates / nodes in a neural network model (\citealp{zhu2022}). Alternatively, spatial autocorrelation can be embedded into the optimization function of the ML model (\citealp{saha2023}, \citealp{Zhan2025}), or a de-correlation preprocessing step can be applied before applying the ML algorithm (\citealp{heaton2024}). However, these approaches are based on ML models and hence have limited inferential capabilities, including not being able to provide relative risk based ERFs with full UQ.

Therefore here we adopt a \emph{residual learning} approach using a stacked modeling strategy, which sequentially applies  ML and spatial smoothing models to the data by using the output from one model as an input to the next model. This leverages the flexibility of ML algorithms for capturing non-linear and interacting confounder-response relationships, with the ability of Bayesian spatial smoothing models for capturing residual autocorrelations and providing interpretable ERFs with full UQ. The first such two-stage approach by \cite{hengl2015} combined a random forest model in stage one with a second stage smoothing model that Kriged the residuals from the random forest. However, this approach applied the random forest to the data assuming independence, which is well known to lead to sub-optimal inference in the presence of spatial autocorrelation. Therefore, \cite{macbride2025} and \cite{figueira2025} recently proposed iterative estimation algorithms that sequentially fit random forest and spatial autocorrelation model components to the data, and terminate based on optimizing the model's out-of-sample predictive performance and the  Kullback-Leibler divergence respectively.

This existing spatial-ML literature focuses almost exclusively on  predicting a spatial process at unmeasured locations, which is not the inferential target in an epidemiological association study. Therefore, this paper proposes the \texttt{SPAR-Forest-ERF} algorithm, which is the first coherent spatial-ML framework designed for estimating a range of ERFs in the presence of known confounders with non-linear and interacting effects and residual spatial autocorrelation. The algorithm extends the iterative residual learning strategy of \cite{macbride2025} through three key developments. First, it introduces a novel stopping rule based on the ERF. Second, it propagates uncertainty between the two model components, yielding more comprehensive uncertainty quantification than \cite{figueira2025}. Finally, it estimates a range of linear and non linear ERFs to maximise model flexibility. This methodology is motivated by a new study of the impact of air pollution concentrations on self-rated health in Scotland, using data from the recently released 2022 national census. These data are described in Section 2, while the methodological development is outlined in Section 3. A simulation study assessing model performance is describe in Section 4, while the motivating study results are presented in Section 5 before a conclusion in Section 6.

\section{Motivating study}
The most recent Scottish census happened in 2022, and is a survey of the entire Scottish population undertaken every 10 years. This study investigates the association between air pollution  and self-rated health in Scotland using these census data, which are available from \url{https://www.scotlandscensus.gov.uk/}. This study is the first of its kind in Scotland following recent studies in Germany (\citealp{liao2025}) and the United Kingdom (\citealp{abed2022}). Scotland is partitioned into $K=6,972$ Data Zones (DZ), which contain 738 people on average (1$st$ percentile - 396, 99$th$ percentile - 2,032) and are the smallest areal unit geography at which the data are publicly available.

\subsection{Disease data}
The disease outcome is self-assessed general health, resulting from the census question \emph{how is your health in general?}. The allowable responses were \emph{Very good}, \emph{Good}, \emph{Fair}, \emph{Bad}, and \emph{Very bad}; and the data comprise counts of the numbers of people who answered in each category for each DZ. Here, we consider two health outcomes to assess the robustness of our conclusions: (i)  the number of people who consider themselves in bad health (answered \emph{Bad} or \emph{Very bad}); and (ii) the number of people who consider themselves in good health (answered \emph{Good} or \emph{Very good}). These DZ level counts will depend on the age-sex demographics of the local populations, which we account for by computing the expected numbers of people in bad / good health in each DZ using indirect standardization. 

The standardized morbidity ratio (SMR) is an exploratory measure of the risk of bad / good health in each DZ, which is computed by dividing the observed count by the corresponding expected count. An SMR of one corresponds to the Scottish average risk, while values respectively above / below one corresponding to elevated / reduced risks compared to this country-wide average. The distributions of these DZ-level SMR values are summarized in Table 1 in the supplementary material, which shows that the distribution for bad health has much more variation than the corresponding distribution for good health, which is due to the latter being based on much larger counts. Finally, to illustrate the spatial patterning in the SMRs, maps of the SMR values for the two largest cities in Scotland (Edinburgh and Glasgow) are also presented in Section 1.1 of the supplementary material.

\subsection{Exposure data}
We focus on nitrogen dioxide (NO$_2$) and particulate matter with an aerodynamic diameter less than or equal to $10 \mu m$ (PM$_{10}$) and $2.5 \mu m$ (PM$_{2.5}$) in this study, because these pollutants are responsible for most of the 26 air quality management areas in Scotland, for details see \url{http://www.scottishairquality.scot/laqm/aqma}. These pollutants are measured by a sparse network of fixed-site monitors, which in 2022 numbered around 85 spatial locations predominantly located in the four major urban centers (see Figure 3 in the supplementary material). As this network has very poor spatial coverage of the $K=6,972$ DZs that make up the country we estimate pollutant levels using estimated annual average concentrations for 2022 from the Pollution Climate Mapping (PCM) model (\url{https://uk-air.defra.gov.uk/data/pcm-data}). The model was developed for the Department for the Environment, Food and Rural Affairs (DEFRA), and is used to monitor concentrations of air pollutants as required by the EU Directive 2008/50/EC. 

The PCM estimates are available on a 1km$^2$ grid, which is spatially misaligned with the irregularly shaped DZs. This spatial change of support is overcome using spatially weighted averaging, where the concentration in each DZ is estimated as the mean value of all the grid square concentrations whose centroids lie within that DZ. For DZs that do not contain any grid square centroids the estimated concentration from the nearest grid square is used instead. The distributions of these DZ level concentrations are summarized in Table 1 in the supplementary material, which shows that pollution levels in Scotland are generally low with median values of 6.69$\mu gm^{-3}$ (NO$_{2}$), 4.66$\mu gm^{-3}$ (PM$_{2.5}$) and 8.73$\mu gm^{-3}$ (PM$_{10}$) respectively. For context, the UK air quality strategy sets objectives for the annual mean concentrations of these pollutants in Scotland to be below 40$\mu gm^{-3}$ (NO$_{2}$), 10$\mu gm^{-3}$ (PM$_{2.5}$) and 18$\mu gm^{-3}$ (PM$_{10}$) respectively. Concentrations of all three pollutants are higher in the urban areas compared to the rural areas as expected, while both PM$_{2.5}$ and PM$_{10}$ show elevated concentrations on the eastern coast due to trans-boundary pollution coming from continental Europe.

\subsection{Confounder data}
The main confounders available in this study are summarized in Table 1 of the supplementary material. The most prominent of these is socio-economic deprivation, which in Scotland is measured by the Scottish index of multiple deprivation (SIMD, \url{https://simd.scot/}). The SIMD is a composite index comprising 27 separate indicators, and those relating to health are removed because modeling health data in terms of health data is circular. Additionally, the indicators relating to crime rates, school attendance and a continuous school attainment score are also removed due to their large numbers of missing values (501, 567 and 189 respectively). Finally, any pair of indicators whose correlation is over 0.85 in absolute value had one variable in that pair removed, resulting in a final set of 12 indicators that are summarized in Table 2 in the supplementary material.

We also obtained data on ethnicity, specifically, the percentages of people in each DZ that come from the following ethnic groupings: (i) White; (ii) Indian / Pakistani / Bangladeshi; (iii) Chinese; (iv) African; and (v) Caribbean or Black. The first of these is removed  due to its collinerity with the remaining variables that arises because their sum is close to 100\% for all DZs.  The next set of covariates concerns how urban or rural a DZ is, which is measured by both population density (number of people per 1km$^2$) and two binary indicator variables denoting whether each DZ is respectively urban (part of a settlement of over 10,000 people) or rural (part of a settlement of less than 3,000 people).  The final confounder measures loneliness, which we approximate using census data quantifying the percentage of people in each DZ who live in single person households.

\subsection{Exploratory analysis}
We examined the relationships between disease risk ($\ln(\mbox{SMR})$) and the covariates using correlation analysis (see Section 1.2 of the supplementary material), and we found that: (i) the pollutants should be included in separate models due to their collinearity; and (ii) there is unlikely to be any remaining collinearity amongst the final set of confounders. We then examined the presence of residual spatial autocorrelation in both health outcomes, by fitting both a simple Poisson log-linear model and a random forest (on the log SMR scale) to the data. The tuning parameters of the random forest were chosen by minimizing the root mean square error in the out-of-bag errors, for more details see Section 3. The presence of spatial autocorrelation is assessed via Moran's I permutation tests (\citealp{moran1950}), which are based on a neighborhood matrix  defined by the border sharing rule and 10,000 random permutations. The observed I statistics are 0.239 (Poisson glm) and 0.110 (random forest) for bad health, and 0.170 (Poisson glm) and 0.110 (random forest) for good health. These results suggest that the random forests have captured more of the spatial autocorrelation in both data sets via the covariates than the simpler log-linear models have. However, all p-values against the null hypothesis of independence are less than 0.0001, suggesting that despite its superior explanatory power the random forest model is not sufficient on its own for capturing the spatial structure in both health outcomes, which motivates its fusion with a spatial smoothing model in the next section.

\section{Methodology}
The spatial autoregressive random forest algorithm for estimating exposure-response functions is denoted by \texttt{SPAR-Forest-ERF}, and fuses random forests for capturing flexible confounder-response associations with Bayesian spatial smoothing models for estimating a range of ERFs with full UQ while allowing for residual spatial autocorrelation. The flexible model class is presented in Section 3.1, while the inferential algorithm is outlined in Section 3.2.

\subsection{SPAR-Forest-ERF model}
Partition the study region into $K$ non-overlapping areal units $\mathcal{S}=\{\mathcal{S}_1,\ldots,\mathcal{S}_K\}$, and let $\mathbf{Y}=[Y(\mathcal{S}_1),\ldots,Y(\mathcal{S}_K)]^{\top}$ and $\mathbf{e}=[e(\mathcal{S}_1),\ldots,e(\mathcal{S}_K)]^{\top}$ respectively denote $K\times 1$ vectors containing the observed and expected disease counts. Additionally, let $\mathbf{x}(\mathcal{S}_k)=[x_1(\mathcal{S}_k),\ldots,x_q(\mathcal{S}_k)]_{1\times q}$ denote the vector of $q$ exposures for areal unit $\mathcal{S}_k$, while $\mathbf{z}(\mathcal{S}_k)=[z_1(\mathcal{S}_k),\ldots,z_p(\mathcal{S}_k)]_{1\times p}$ denotes a vector of $p$ confounders. Here, the effects of the exposures on disease risk are the inferential target, while the confounders represent nuisance covariates not of direct interest.  The proposed SPAR-Forest-ERF model has the general form:

\begin{eqnarray}
   Y(\mathcal{S}_k)&\sim&\mbox{Poisson}[e(\mathcal{S}_k)\theta(\mathcal{S}_k)]~~~~\mbox{for }k=1,\ldots,K\label{eq:like}\\
    \ln[\theta(\mathcal{S}_k)]&=&  \beta_0 + g[\mathbf{x}(\mathcal{S}_k)] + m[\mathbf{z}(\mathcal{S}_k)] + \phi(\mathcal{S}_k),\nonumber
\end{eqnarray}

where $\theta(\mathcal{S}_k)$ denotes disease risk in $\mathcal{S}_k$ relative to $e(\mathcal{S}_k)$. The spatial pattern in the log risk surface is modeled by the exposures of interest $\{\mathbf{x}(\mathcal{S}_k)\}$, the confounders $\{\mathbf{z}(\mathcal{S}_k)\}$ and a set of random effects $\{\phi(\mathcal{S}_k)\}$ that account for any residual spatial autocorrelation, while the weakly informative prior $\beta_0\sim\mbox{N}(0, 100000)$ is assigned to the intercept term.

\subsubsection{Exposure-response function $g[\mathbf{x}(\mathcal{S}_k)]$}
A range of functional forms can be specified for the ERF $g[\mathbf{x}(\mathcal{S}_k)]$ depending on the goal of the analysis, and in what follows we outline three possible options in the case of a single exposure, i.e.,
$\mathbf{x}(\mathcal{S}_k) =x(\mathcal{S}_k)$ where $q=1$.

\begin{enumerate}
    \item[I] \texttt{Linear ERF} - $g[x(\mathcal{S}_k)]=\alpha x(\mathcal{S}_k)$, which is the simplest specification and allows one to quantify the overall linear association $\alpha$ between the exposure and the risk of disease. 
    \item[II] \texttt{Non-linear ERF} -  $g[x(\mathcal{S}_k)]$, which allows the association to be a smooth non-linear function whose size depends on the level of the exposure. This smooth function is represented by a Bayesian p-spline, and is fitted as a second order random walk latent field model using the INLA (\citealp{rue2009}) software, following the approach outlined in \cite{wang2018}.
    \item[III] \texttt{Measurement error ERF} - $g[x(\mathcal{S}_k)]=\alpha x(\mathcal{S}_k)$ and $x(\mathcal{S}_k)\sim\mbox{N}(\tilde{x}(\mathcal{S}_k), \sigma^2_x)$, which uses a Berkson measurement error model to relate the true unknown exposure $x(\mathcal{S}_k)$ to the available error-prone estimate $\tilde{x}(\mathcal{S}_k)$. In the motivating application the error variance $\sigma^2_x$ can be estimated from the modeled and measured pollution data as described in Section 2.1 of the supplementary material.
\end{enumerate} 

For options I and III we assume the weakly informative prior   $\alpha\sim\mbox{N}(0, 100000)$.

\subsubsection{Confounder component $m[\mathbf{z}(\mathcal{S}_k)]$}
The associations between the confounders and the natural log of disease risk $m[\mathbf{z}(\mathcal{S}_k)]$ is estimated by a random forest (RF), because it is fast to implement, can capture non-linear and interacting confounder-response association, whilst producing approximately out-of-sample predictions for the data via their bootstrapping construction using out-of-bag (OOB) predictions. Additionally, existing research shows they often outperform neural networks for tabular (\citealp{Grinsztajn2022}) and spatial (\citealp{macbride2025}) data, whilst generalized additive models require the specific interactions between confounders to be specified in advance and not learned from the data. A brief summary of random forests and their tuning parameters is given in Section 2.2 of the supplementary material.

\subsubsection{Random effects component $\phi(\mathcal{S}_k)$}
The $K\times 1$ vector of random effects $\boldsymbol{\phi}=[\phi(\mathcal{S}_1),\ldots,\phi(\mathcal{S}_K)]^{\top}$ captures the remaining spatial autocorrelation in disease risk after covariate adjustment, as well as any overdispersion with respect to the Poisson data likelihood. They are commonly assigned a conditional autoregressive (CAR) prior distribution, whose spatial autocorrelation structure depends on a $K\times K$ neighborhood matrix $\mathbf{W}=(w_{kj})$, where $w_{kj}=1$ if areal units $(\mathcal{S}_k, \mathcal{S}_j)$ share a common border and $w_{kj}=0$ otherwise (with $w_{kk}=0~\forall k$). Any areal unit that does not have any neighbors is assigned one neighbor based on its smallest inter-centrodial distance with the remaining areal units. A number of different CAR priors have been proposed and could be incorporated into SPAR-Forest-ERF, and here we use the modified Besag-York-Mollie model proposed by \cite{riebler2016} because it contains a set of random effects $\mathbf{u}=[u(\mathcal{S}_1),\ldots,u(\mathcal{S}_K)]^{\top}$ that capture spatial autocorrelation and a second independent set $\mathbf{v}=[v(\mathcal{S}_1),\ldots,v(\mathcal{S}_K)]^{\top}$ that captures  overdispersion as follows.

\begin{eqnarray}
    \boldsymbol{\phi}&=& \frac{1}{\sqrt{\tau_{\phi}}}\left[\sqrt{(1-\rho)}\mathbf{v} +  \sqrt{\rho}\mathbf{u}\right]\label{bym2}\\
    \mathbf{v}&\sim&\mbox{N}(\mathbf{0}, \mathbf{I})\nonumber\\
    \mathbf{u}&\sim&\mbox{N}(\mathbf{0}, \tilde{\mathbf{Q}}(\mathbf{W})^{-})\nonumber.
    \end{eqnarray}

Here,  $\mathbf{v}$ are modeled as independent zero-mean Gaussian random variables with unit variance, with the amount of variation in $\boldsymbol{\phi}$ controlled by the global precision  parameter $\tau_{\phi}$. The proportion of this variation that is spatially autocorrelated is controlled by $\rho$, with $\rho=0$ corresponding to independence while $\rho=1$ corresponds to strong spatial autocorrelation. The spatially autocorrelated random effects  $\mathbf{u}$ are modeled with a zero-mean multivariate Gaussian distribution, whose singular precision matrix $\tilde{\mathbf{Q}}(\mathbf{W})$ (here $^-$ denotes a generalized inverse) is a scaled version of  the precision matrix from the intrinsic CAR model that ensures the geometric mean of the marginal variances equals one. Finally, the precision parameter $\tau_{\phi}$ is assigned a half normal prior on the standard deviation scale with precision 0.001 (i.e., $\tau_{\phi}^{-1/2}\sim\mbox{Half-Normal}(\tau_0=0.001)$), while $\rho$ is assigned a penalized complexity prior on the $\ln(\rho/(1-\rho))$ scale because it was suggested by \cite{riebler2016}.

\subsection{Model fitting and inference}
Model (\ref{eq:like}) is fitted using a novel iterative two-step residual learning strategy, which extends   \cite{macbride2025} and \cite{figueira2025} by: (i) allowing for full uncertainty propagation between both steps of the algorithm; (ii) proposing a new stopping criteria based on the ERF (the primary inferential target); and (iii) producing inference for the ERF with full UQ via its posterior distribution. The fitting process is outlined in \texttt{Algorithm 1} below, and is initialized by setting $g[\mathbf{x}(\mathcal{S}_k)]=\phi(\mathcal{A}_k)=0~~\forall k$. It then iterates between estimating $\{m[\mathbf{z}(\mathcal{S}_k)]\}$ using a random forest (\textbf{Step 1}) and $\{g[\mathbf{x}(\mathcal{S}_k)], \phi(\mathcal{A}_k)\}$ using a Bayesian spatial smoothing model (\textbf{Step 2}), where each component is fitted after adjusting for the current values and associated uncertainties of the other component. Note, to speed up the algorithm no uncertainty is allowed for when selecting the optimal values of the random forest tuning parameters. The algorithm terminates when the mean absolute difference in $\{g[\mathbf{x}(\mathcal{S}_k)]\}$ between successive iterations is less than  $\epsilon=0.0005$, which simulation studies showed  leads to good model performance. 

In \textbf{step 1} $\{m[\mathbf{z}(\mathcal{S}_k)]\}$ is estimated using OOB predictions $\{\hat{m}[\mathbf{z}(\mathcal{S}_k)]\}$, and we propagate the uncertainties in these predictions  by  adding  a Berkson measurement error component to the Bayesian spatial smoothing model, because the true values $\{m[\mathbf{z}(\mathcal{S}_k)]\}$ are likely to have additional random variation compared to the smoothed OOB estimates $\{\hat{m}[\mathbf{z}(\mathcal{S}_k)]\}$. This yields the  \textbf{Step 2} model:

\begin{eqnarray}
Y(\mathcal{S}_k)&\sim&\mbox{Poisson}[e(\mathcal{S}_k)\theta(\mathcal{S}_k)]~~~~\mbox{for }k=1,\ldots,K\label{eq:like2}\\
    \ln[\theta(\mathcal{S}_k)]&=&  \beta_0 + g[\mathbf{x}(\mathcal{S}_k)] +  m[\mathbf{z}(\mathcal{S}_k)] + \phi(\mathcal{S}_k)\nonumber\\
    m[\mathbf{z}(\mathcal{S}_k)] &=& \hat{m}[\mathbf{z}(\mathcal{S}_k)] + \varepsilon(\mathcal{S}_k),~~~~\varepsilon(\mathcal{S}_k) \sim \mbox{N}(0, \hat{\sigma}^2_{m}),\nonumber
\end{eqnarray}

where the remaining model components are as defined as in Section 3.1. The variance of the prediction errors $\{\varepsilon(\mathcal{S}_k)=m[\mathbf{z}(\mathcal{S}_k)]-\hat{m}[\mathbf{z}(\mathcal{S}_k)]\}$ is denoted by $\hat{\sigma}^2_{m}$, and following \cite{figueira2025} is estimated by the root mean square error of the out-of-bag prediction errors. 

Unlike \cite{figueira2025} and \cite{macbride2025}, we also incorporate the posterior uncertainty from  (\ref{eq:like2}) into the next iteration of random forest in \textbf{Step 1}. This is achieved by generating $q=1,\ldots Q=100$  posterior samples for $\{g[\mathbf{x}(\mathcal{S}_k)]_q +  \phi(\mathcal{S}_k)_q\}_{q=1}^{Q}$ from  (\ref{eq:like2}), which are used to compute $Q$ samples from the set of adjusted response variables $\{R(\mathcal{S}_k)_{q} = \ln[Y(\mathcal{S}_k)/e(\mathcal{S}_k)] - g[\mathbf{x}(\mathcal{S}_k)]_q - \phi(\mathcal{S}_k)_q\}_{k=1}^{K}$ used in the next iteration of the random forest. These adjusted responses correspond to the linear predictor scale of (\ref{eq:like2}) adjusted for the current values of the ERF and the random effects, which is the scale at which the confounders enter the model. A random forest of 10 trees is then fitted to each of these $Q$ sets of response variables, and these 100 random forests are combined to create an overall random forest of 1,000 trees that incorporates the posterior uncertainty in $\{g[\mathbf{x}(\mathcal{S}_k)] +  \phi(\mathcal{S}_k)\}_{k=1}^{K}$. The OOB predictions from this random forest are the mean predictions from the sub-forest of trees that were fitted without the relevant data point in question, which together with its error variance are computed by

$$\hat{m}[\mathbf{z}(\mathcal{S}_k)]=\frac{1}{\sum_{q=1}^{Q}n_{qk}}\sum_{q=1}^{Q}n_{qk}\tilde{m}[\mathbf{z}(\mathcal{S}_k)]_q~~~~\mbox{and}~~~~
\hat{\sigma}^2_m=\frac{1}{K}\sum_{k=1}^{K}(\tilde{R}(\mathcal{S}_k)- \hat{m}[\mathbf{z}(\mathcal{S}_k)])^2.$$

Here, $\{\tilde{m}[\mathbf{z}(\mathcal{S}_k)]_q, n_{qk}\}$ respectively denote the OOB prediction and the number of trees upon which that prediction was based from the random forest based on the $q$th data sample for the $k$th observation. The \emph{average} response variable is similarly computed as $\tilde{R}(\mathcal{S}_k)=\frac{1}{\sum_{q=1}^{Q}n_{qk}}\sum_{q=1}^{Q}n_{qk}R(\mathcal{S}_k)_q$.

\begin{algorithm}
\caption{\texttt{SPAR-Forest-ERF}}
\label{alg:spar_forest}
\begin{algorithmic}[1] % [1] for line numbering
    \Statex \textbf{Initialization}
    \State Initialize  $g[\mathbf{x}(\mathcal{S}_k)]^{(0)}=\phi(\mathcal{A}_k)^{(0)}=0~~\forall k$, and choose the stopping threshold $\epsilon$.
   \vspace{0.2cm}

    \Statex \textbf{Iterate} 
    \State Iterate \textbf{Step 1} and \textbf{Step 2} over $i=1,2,3,\ldots,$ until $\frac{1}{K}\sum_{k=1}^{K}\left|g[\mathbf{x}(\mathcal{S}_k)]^{(i)}-g[\mathbf{x}(\mathcal{S}_k)]^{(i-1)}\right| < \epsilon$.
        
    \State \hspace{\algorithmicindent}\textbf{Step 1.} Generate $q=1,\ldots,Q$ samples from the set of adjusted response variables $\{R(\mathcal{S}_k)^{(i)}_q = \ln[Y(\mathcal{S}_k)/e(\mathcal{S}_k)] - g[\mathbf{x}(\mathcal{S}_k)]^{(i-1)}_q - \phi(\mathcal{S}_k)^{(i-1)}_q\}_{k=1}^{K}$, where if $i=1$ then $g[\mathbf{x}(\mathcal{S}_k)]^{(i-1)}_q = \phi(\mathcal{S}_k)^{(i-1)}_q=0$ for all $q$. Tune a random forest  with covariates $\{\mathbf{z}(\mathcal{S}_k)\}$ and response variables that are the  mean (taken over $q=1,\ldots,Q$) values of the response variables $\{R(\mathcal{S}_k)^{(i)}_q\}$ to obtain optimal values of \texttt{mtry} and \texttt{minnode}. Based on these optimal values fit separate random forests for $q=1,\ldots,Q$ samples of the response variables, and construct the out-of-bag predictions $\{\hat{m}[\mathbf{z}(\mathcal{S}_k)]^{(i)}\}$ and their variance $\hat{\sigma}^{2^{(i)}}_m$.
    
    \State \hspace{\algorithmicindent}  \textbf{Step 2.} Fit the Bayesian spatial smoothing model (\ref{eq:like2}) using INLA, where the true confounder component  $\{m[\mathbf{z}(\mathcal{S}_k)]\}_{k=1}^{K}$ is represented by a Berkson measurement error model based on the estimates $\{\hat{m}[\mathbf{z}(\mathcal{S}_k)]^{(i)}\}_{k=1}^{K}$ and the error variance $\hat{\sigma}^{2^{(i)}}_m$ computed in \textbf{Step 1}. Generate $q=1,\ldots,Q$ samples from the posterior distribution of  $\{g[\mathbf{x}(\mathcal{S}_k)]^{(i)}_q +  \phi(\mathcal{A}_k)^{(i)}_q\}_{k=1}^{K}$ from this model.
    \Statex \textbf{End iterate}\vspace{0.2cm}

 \Statex \textbf{Inference}
 \State Inference for the exposure response function $\{g[\mathbf{x}(\mathcal{S}_k)]\}_{k=1}^{K}$  and the random effects $\{\phi(\mathcal{A}_k)\}_{k=1}^{K}$ are obtained from their posterior distributions from the final Bayesian spatial smoothing model fitted in \textbf{Step 2}, while inference about the confounders is given by the final out-of-bag estimate $\{\hat{m}[\mathbf{z}(\mathcal{S}_k)]^{(i)}\}_{k=1}^{K}$  obtained from \textbf{Step 1.}
\end{algorithmic}
\end{algorithm}

\subsection{Software}
Software to implement the SPAR-Forest-ERF algorithm is freely available from \url{https://github.com/vinnydavies/SPARforest} with a accompanying documentation. It is coded in \texttt{R}, and the random forest component is fitted using a recursive binary partitioning algorithm via the \texttt{ranger} (\citealp{wright2017}) package, while the hierarchical Bayesian model is fitted using integrated nested Laplace approximations with variational Bayes approximations via the \texttt{INLA} (\citealp{rue2009}) package.

\section{Simulation study}
This study quantifies the accuracy with which the proposed SPAR-Forest-ERF algorithm and its competitors can estimate ERFs under a range of different conditions, including different disease prevalences; ERF shapes, types of confounding, and levels of residual spatial autocorrelation.

\subsection{Data generation}
The study is based on the set of $K=3,066$ Data Zones comprising the central belt of Scotland including Glasgow and Edinburgh, with the whole of Scotland not used because its large data  size makes the repeated model fitting computationally prohibitive for hundreds of simulated data sets. Disease count data are generated from independent Poisson distributions with means $\{e(\mathcal{S}_k)\theta(\mathcal{S}_k)\}$, where the expected counts $\{e(\mathcal{S}_k)\}$ are varied across the scenarios to determine what impact disease prevalence has on estimation accuracy. The spatial risk surface $\{\theta(\mathcal{S}_k)\}$ is generated on the natural log-scale as a function of exposure variables, known confounders and a set of random effects, the latter representing the  residual spatial pattern in disease risk not captured by the available covariates. Two exposure variables $\{x_1(\mathcal{S}_k),x_2(\mathcal{S}_k)\}$ with different levels of spatial autocorrelation are constructed for each simulated data set, to see what impact their dependence structure has on the estimation of their ERFs. The first is generated independently over space from a standard normal distribution. The second is assumed to be spatially autocorrelated to mimic air pollution concentrations, and is drawn from a zero-mean multivariate normal distribution. Its covariance matrix is specified by an exponential autocovariance function based on the pairwise distances between DZ centroids, where the range parameter is chosen so that the correlation is 0.75 for locations 4.45km apart (the 5$th$ percentile of the distribution of inter-centroidal distances) to ensure strong spatial autocorrelation. Both exposure variables are then rescaled to the unit interval to ensure they have the same positive exposure range. 

Eight separate confounders are generated for each simulated data set, not all of which contribute to the true risk surface. The first four are independent in space, and are generated from standard normal distributions as above before being rescaled to the unit interval. The latter four are spatially autocorrelated and generated from a zero-mean multivariate normal distribution with a spherical autocorrelation function that is zero for DZs more than 5km apart, before again being rescaled to the unit interval. These confounders are related to disease risk via
$m[\mathbf{z}(\mathcal{S}_k)]= 3 / \{1 + \exp[6 - 12z_1(\mathcal{S}_k)]\} + 2[z_3(\mathcal{S}_k)z_6(\mathcal{S}_k)-0.4]^2 +  \cos[2\pi z_7(\mathcal{S}_k)] + z_8(\mathcal{S}_k)$, which includes linear and non-linear relationships as well as interactions. Finally, the residual spatial autocorrelation is generated from another zero-mean multivariate normal distribution, whose covariance matrix corresponds to the CAR model proposed by \cite{leroux2000} with a spatial dependence parameter fixed at $\rho=0.95$ to induce relatively strong autocorrelation.

\subsection{Study design}
One hundred simulated data sets are generated under each of  9 different scenarios, the first 8 of which collectively cover all possible combinations of the three factors listed below. Additionally, the ninth scenario has exactly linear confounder-response associations with no interactions via $m[\mathbf{z}(\mathcal{S}_k)]=0.5 z_1(\mathcal{S}_k) - z_3(\mathcal{S}_k) + 2z_7(\mathcal{S}_k) - 0.1z_8(\mathcal{S}_k)$, which allows us to  observe how well SPAR-Forest-ERF does in this simplified situation. 

\begin{itemize}
    \item \textbf{Disease prevalence} - (A) \texttt{Rare} - where $e(\mathcal{S}_k)\sim \mbox{Uniform}[10, 25]$; and (B) \texttt{Common} - where  $e(\mathcal{S}_k)\sim \mbox{Uniform}[100, 200]$. This allows us to assess how well the models perform when there are different amounts of information in the data. 

    \item \textbf{True ERF shapes} - (A) \texttt{Linear} - $g[x_j(\mathcal{S}_k)]=x_j(\mathcal{S}_k)\alpha$; and (B) \texttt{Sigmoidal} -  $g[x_j(\mathcal{S}_k)]=\alpha / [1 +  \exp(6-12 x_j(\mathcal{S}_k))]$, for each covariate $x_j(\mathcal{S}_k)$ for $j=1,2$. The shapes of these ERFs are displayed in Figure  5 in the supplementary material, and while a linear shape is commonly assumed in epidemiological air pollution studies for simplicity, a sigmoidal shape has been found in a recent harmonizing analysis of 3 large cohort studies (\citealp{chen2023}). Here we fix $\alpha=0.2$ because this gives the ERFs a similar range to that  observed for PM$_{2.5}$ in the motivating study.
    
    \item \textbf{Control for confounding} - (A) \texttt{Good} -  $m[\mathbf{z}(\mathcal{S}_k)]$ captures 95\% of the variation in the data not explained by the exposures, so that the unmeasured confounding represented by $\phi(\mathcal{S}_k)$ only accounts for 5\% of this variation; (B) \texttt{Poor} -  $m[\mathbf{z}(\mathcal{S}_k)]$ captures 60\% of the variation in the data not explained by the exposures, so that the unmeasured confounding represented by $\phi(\mathcal{S}_k)$ accounts for 40\% of this variation.
\end{itemize}    

Three different models are compared in this study, the first of which is the standard spatial epidemiological risk model denoted by \texttt{GLMM} that is a simplification of (\ref{eq:like}) with linear confounder-response associations, i.e., $m[\mathbf{z}(\mathcal{S}_k)]=\mathbf{z}(\mathcal{S}_k)^{\top}\boldsymbol{\delta}$.  The second is the SPAR-Forest-ERF algorithm proposed here (denoted \texttt{SPAR-Forest-ERF}), while the third is a simplification of it that is non-iterative (each of the two model components are only fitted once) and does not propagate uncertainty between the two model components (denoted \texttt{SPAR-Forest-ERF-1}). This simplified model is included to determine whether the complexity of the full iterative algorithm is necessary. The relative performances of these models are measured by the accuracy with which they can estimate the exposure response functions $\{g[x_j(\mathcal{S}_k)]\}$ for both the spatially autocorrelated and independent exposures. Let $\{g[x_j(\mathcal{S}_k)], \hat{g}[x_j(\mathcal{S}_k)]\}_{k=1}^{K}$ respectively denote the true and estimated (posterior mean) ERFs, while pointwise 95\% credible intervals are also computed. The accuracy of the estimation from each model is quantified by calculating the following metrics for each of the 100 simulated data sets: (i) \texttt{Bias} $=\frac{1}{K}\sum_{k=1}^{K} \{\hat{g}[x_j(\mathcal{S}_k)] - g[x_j(\mathcal{S}_k)]\}$; (ii) Root mean square error (RMSE) $= \sqrt{\frac{1}{K}\sum_{k=1}^{K} \{\hat{g}[x_j(\mathcal{S}_k)] - g[x_j(\mathcal{S}_k)]\}^2}$; (iii) coverage $=$ percentage of the pointwise 95\% credible intervals for $\{g[x_j(\mathcal{S}_k)]\}$ that contain the true value; and (iv) mean interval width (MIW) $=$ mean width of the pointwise 95\% credible intervals for $\{g[x_j(\mathcal{S}_k)]\}$. The bias summarizes whether each model over or under estimates the ERF on average, while the RMSE quantifies the accuracy of the point estimation. Finally, the coverage and MIW metrics summarize the accuracy with which the models quantify the uncertainty in these estimates.

\subsection{Results}
The results of the study are presented below and in Section 3.3 of the supplementary material, while a summary of the number of iterations taken by the SPAR-Forest-ERF algorithm and the mean absolute difference between the estimated ERFs from its final two iterations are presented in Section 3.2 of the supplementary material. All three models produce essentially unbiased estimates of the ERFs in all scenarios over all 100 simulated data sets, and these results are hence not shown for brevity. Briefly, these biases range between (-0.006, 0.006) for the autocorrelated exposure and between (-0.001, 0.001) for the independent exposure, which applies regardless of whether the ERF is linear or sigmoidal. 

\begin{figure}
    \centering
    \includegraphics[width=1\linewidth]{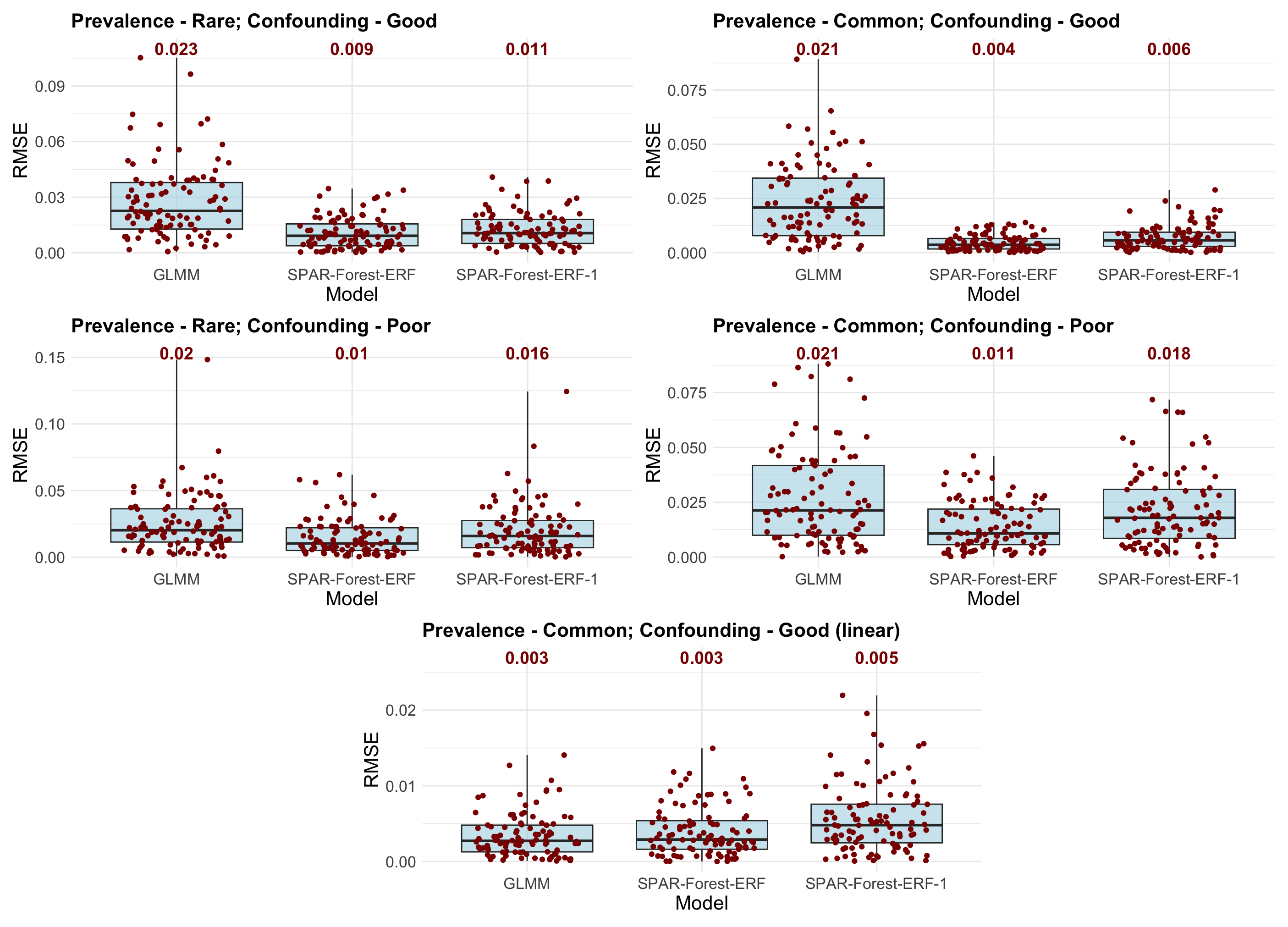}
    \caption{Root mean square errors (boxplots and points) for the estimates of the linear ERFs corresponding to the spatially autocorrelated exposure. The top four panels relate to different disease prevalences (rare vs common) and control for confounding (good vs poor), while the bottom panel relates to where all the known confounders have simple linear relationships with disease risk. The numbers relate to the mean RMSE over the 100 simulated data sets.}
    \label{fig:rmsecorrlinear}
\end{figure}

The root mean square errors for the estimates of the linear ERF corresponding to the spatially autocorrelated exposure are displayed in Figure \ref{fig:rmsecorrlinear} for each of the three models, and we focus on this exposure in the main paper because it is the one that mimics air pollution in the motivating study. The top four panels relate to all pairwise combinations of rare vs common diseases and good vs poor control for confounding, while the bottom panel relates to the simplistic setting where all the known confounders have exactly linear relationships with disease risk. The set of RMSEs across the 100 simulated data sets are displayed using both boxplots and the actual values (jittered horizontally) for clarity, while the median RMSE in each scenario is given above each boxplot. The Figure shows that the SPAR-Forest-ERF algorithm greatly outperforms the commonly used GLMM in all cases where the confounder-response associations are not exactly linear (top four panels), with percentage reductions in the median RMSE ranging between 49\% and 83\% across the four scenarios. The simplified SPAR-Forest-ERF-1 algorithm has RMSEs between those from the other two models, with percentage reductions compared to the GLMM (in the median values) ranging between 16\% and 73\%. Finally, in the simplistic scenario where all the confounders have exactly linear relationships with disease risk the GLMM and SPAR-Forest-ERF models produce comparable results, even though the data generating mechanism matches and hence favors the GLMM. The corresponding RMSE results for the independent exposure and for both exposures where the true ERFs are sigmoidal are presented in Section 3.3 of the supplementary material. 

The pointwise percentage coverages and mean interval widths of the 95\% credible intervals for the spatially autocorrelated exposure with a linear ERF are displayed in Table \ref{tab:coveragespatiallinear}, where the results have been averaged (mean) over all 100 simulated data sets in each scenario. Note, as the true ERF is linear then each 95\% credible interval either does or does not contain the true ERF for all $K$ observations, which results in integer coverages when averaged over the 100 simulated data sets. The remaining results concerning coverages and mean interval widths are presented in Section 3.3 of the supplementary material. The table shows that the SPAR-Forest-ERF algorithm has coverages close to the nominal 95\% level  in almost all cases (range - 80\% - 99\%), while the corresponding coverages for the GLMM are much lower ranging between 56\% and 94\%. The simplified SPAR-Forest-ERF-1 algorithm has coverages that mostly sit between those from the other two models, with a range between 63\% and 94\%. As well as having improved coverages the SPAR-Forest-ERF algorithm also has more precise credible intervals compared to those from the GLMM, because their mean widths are narrower by between 9\% and 52\% across the 8 scenarios with non-linear confounding. In the linear confounding scenario the intervals are slightly wider at 0.023 compared with 0.019. The coverages are lowest for all three models when the spatial random effects represent a sizeable proportion of the variation in disease risk (i.e., when $\{\phi(\mathcal{S}_k)\}$ accounts for 40\% of the confounding) and when the disease is common, which results in coverages as low as 80\% for SPAR-Forest-ERF and 61\% for GLMM.  These low coverages occur because in this setting: (i) the large numbers of disease cases means there is a lot of information in the data resulting in relatively narrow credible intervals; and (ii) the spatially autocorrelated random effects are heavily influential and confound the estimated effect of the spatially smooth exposure (see \citealp{hodges2010}) leading to relatively poor estimation.

\begin{table}
    \centering
    \begin{tabular}{llrrr}\toprule
\textbf{Prevalence}   &\textbf{Confounding}  &\textbf{GLMM}  &\textbf{SPAR-Forest-ERF}  &\textbf{SPAR-Forest-ERF-1} \\\midrule
     \underline{Coverage}    &&&&\\
     Rare  & Good  &84 & 99 & 94\\
     Common  & Good  &56 & 98 & 84\\
     Rare & Poor & 83 & 93 & 88\\
    Common  & Poor & 61 & 80 & 63\\
     Common &  Good (linear)  &94 & 99 & 91\\     \midrule
      \underline{MIW}    &&&&\\
     Rare  & Good  &0.090 & 0.069 & 0.054\\
     Common  & Good  &0.048 & 0.023 & 0.024\\
     Rare & Poor &0.086 & 0.068 & 0.075\\
     Common  & Poor &0.050 & 0.045 & 0.045\\
     Common & Good (linear)  &0.019 & 0.023 & 0.022\\
     \bottomrule
    \end{tabular}
    \caption{Coverages (\%) and mean interval widths (MIW) for the linear ERF corresponding to the spatially autocorrelated exposure.}
    \label{tab:coveragespatiallinear}
\end{table}

\section{Results of the motivating study}
The same three models compared in the simulation study are considered here, and each are  applied to the two outcomes of bad and good self-rated health separately using the set of confounders described in Section 2. The three pollutants NO$_{2}$, PM$_{2.5}$ and PM$_{10}$ are included in separate models because it is the predominant approach taken in the epidemiological literature so as to avoid potential issues of collinearity. For each pollutant-outcome pair and model we estimate three different ERFs, including: (I) a simple linear ERF yielding a single interpretable relative risk; (II) a non-linear ERF to allow its true shape to be identified; and (III) a linear ERF allowing for the likely error in the modeled concentrations via a Berkson measurement error model. The results of the linear (I) and non-linear (II) ERFs are presented in Sections 5.2 and 5.3 below, while the measurement error model results (III) are presented in Section 4.1 of the supplementary material. 

\subsection{Model comparison}
The performance of each model with a linear ERF (case I) is summarized in Table \ref{table_modelcompar} by its fit to the observed data, its predictive performance and its ability at capturing the spatial autocorrelation in the data. The fit to the observed data is measured by the deviance $D(\boldsymbol{\Theta})=-2\ln[l(\boldsymbol{\Theta}|\mathbf{Y})]$ (where 
$\boldsymbol{\Theta}$ denotes the model parameters), which is evaluated by both the mean of the posterior distribution of the deviance, $\hat{D}=\mathbf{E}_{\boldsymbol{\Theta}|\mathbf{Y}}[D(\boldsymbol{\Theta})])$, and the deviance evaluated at the posterior mean of the parameters $\hat{\boldsymbol{\Theta}}$ (denoted by $D_{\hat{\boldsymbol{\Theta}}}$). The table shows that  the SPAR-Forest-ERF algorithm provides the best fit to both health outcomes, having the lowest values under both deviance metrics. SPAR-Forest-ERF also exhibits the best cross validated predictive performance for the bad health outcome, because it has the lowest widely applicable information criterion (WAIC) value which is asymptotically equivalent to the Bayes cross-validation loss (\citealp{watanabe2010}). It does however have the worst (highest) WAIC value for the good health outcome, which is due to its much larger estimated number of independent parameters $p_w$. Finally, all models are able to capture the spatial autocorrelation in the data relatively well for both health outcomes, because the Moran's I values for the residuals are close to zero and range between -0.035 and 0.063. This compares to values of 0.239 (bad health) and  0.170 (good health) for the residuals from a simple covariate-only Poisson log-linear model and 0.550 (bad health) and 0.551 (good health) for the raw SMR values.

\begin{table}
    \centering
    \begin{tabular}{lllll}\toprule
     \textbf{Model}   &\textbf{Deviance - $\hat{D}$}  &  \textbf{Deviance -$D_{\hat{\boldsymbol{\Theta}}}$}  & \textbf{WAIC ($p_w$)} &\textbf{Moran's I} \\\midrule

    \texttt{Bad self-rated health}&&&&\\ 
    GLMM &46,885 &43,021 &50,528 (2,767) &-0.0339\\ 
    SPAR-Forest-ERF &45,926 &41,092 &49,464 (2,634)  &0.0320\\ 
    SPAR-Forest-ERF-1 &46,915 &43,558 &50,224 (2,553) &-0.0205\\\midrule
        \texttt{Good self-rated health}&&&&\\ 
    GLMM &62,030 &61,560 &62,369 (315) &0.0289\\ 
    SPAR-Forest-ERF &61,606 &58,732 &63,170 (1,244) &0.0626\\ 
    SPAR-Forest-ERF-1 &61,824 &61,619 &61,964 (136) &0.0224\\\bottomrule
    \end{tabular}
    \caption{Comparison of the relative performance of each model for each health outcome. The model fit measures include: (i) fit to the observed data via the posterior mean deviance ($\hat{D}$) and the deviance evaluated at the posterior mean of the parameters  $\hat{\boldsymbol{\Theta}}$ ($D_{\hat{\boldsymbol{\Theta}}}$); (ii) predictive performance via the WAIC and its effective number of independent parameters ($p_w$); and its ability at capturing spatial autocorrelation via the Moran's I statistic of its residuals.}
    \label{table_modelcompar}
\end{table}

\subsection{Liner exposure response functions - $g[x(\mathcal{S}_k)]$}
A simple linear exposure response function is the standard approach in the epidemiological literature, because it represents the pollution-health association with a single interpretable relative risk (RR). These relative risks quantify the increased risk of disease that is associated with a fixed increase in the yearly average concentration of each pollutant, and here we present RRs relating to the following realistic (close to their spatial standard deviations) increases in each pollutant: NO$_{2}$ - 4$\mu gm^{-3}$; PM$_{2.5}$ - 1$\mu gm^{-3}$; PM$_{10}$ - 2$\mu gm^{-3}$. Estimates and 95\% credible intervals for these RRs are displayed in Table \ref{table_RR}, where the top panel relates to bad self-rated health while the bottom panel relates to good self-rated health.  

The results from the GLMM and SPAR-Forest-ERF models consistently show that particulate matter air pollution (PM$_{2.5}$ and PM$_{10}$) is significantly harmful to self-rated health at the 5\% level, because none of the 95\% credible intervals contain the null RR of one. The estimated relative risks correspond to around a 4\% (PM$_{2.5}$) and a 2.7\% (PM$_{10}$) increased risk of bad self-rated health if particulate matter concentrations increase. The estimated reductions in good self-rated health are much smaller, with reduced relative risks of around 0.4\% for both particulate matter pollutants. Although the point estimates (posterior means) of the relative risks are similar between the GLMM and SPAR-Forest-ERF models, the latter show increased precision with narrower 95\% credible intervals by between 0\% and 24\% for the particulate matter outcomes. In contrast, NO$_{2}$ shows mixed results, with only one of the four relative risks being significant at the 5\% level.

The one iteration simplification of the SPAR-Forest-ERF algorithm leads to attenuated non-significant relative risks close to one in all cases, which is  likely to be because in this algorithm the confounders explain as much of the variation as possible in the risk of disease via the highly flexible random forest component, leaving relatively little variation left to be explained by the exposure (and the random effects). This hypothesis is corroborated by the effective number of independent parameters $p_w$  presented in Table \ref{table_modelcompar}, which are the lowest for the simplified SPAR-Forest-ERF algorithm. Finally, to give an increased visibility to the workings of the SPAR-Forest-ERF algorithm, the evolution of the estimated linear ERFs  across the iterations of the algorithm are presented in Section 4.2 of the supplementary material. 

\begin{table}
    \centering
    \begin{tabular}{llll}\toprule
     \multirow{2}{*}{\textbf{Model}}& \multicolumn{3}{c}{\textbf{Relative risks (RR)}}\\
       &\textbf{NO$_{2}$}  &  \textbf{PM$_{2.5}$}  & \textbf{PM$_{10}$} \\\midrule
        \texttt{Bad self-rated health}&&&\\ 
        GLMM  &1.005 (0.988, 1.022) &1.040 (1.023, 1.057) &1.027      (1.013, 1.041) \\
        SPAR-Forest-ERF  &1.032 (1.021, 1.043) &1.041      (1.028, 1.054) &1.027 (1.015, 1.039) \\
        SPAR-Forest-ERF-1  &1.002      (0.991, 1.012) &1.002 (0.991, 1.013)  &1.007 (0.997, 1.018) \\\midrule
        \texttt{Good self-rated health}&&&\\ 
        GLMM  &1.000      (0.998, 1.003) &0.996      (0.993,      0.999) &0.996      (0.994,      0.998) \\
        SPAR-Forest-ERF  &0.998      (0.996,  1.000) &0.996      (0.993, 0.998) &0.997      (0.995,  0.999) \\
        SPAR-Forest-ERF-1  &0.999      (0.998,  1.001) &1.000      (0.998,   1.002) &0.999      (0.997, 1.001)  \\\bottomrule
    \end{tabular}
    \caption{Summary of the linear exposure-response functions for each pollutant-outcome pair on the relative risk (RR) scale. The relative risks relate to the following increases in each pollutant: NO$_{2}$ - 4$\mu gm^{-3}$; PM$_{2.5}$ - 1$\mu gm^{-3}$; PM$_{10}$ - 2$\mu gm^{-3}$.}
    \label{table_RR}
\end{table}

\subsection{Non-liner exposure response functions - $g[x(\mathcal{S}_k)]$}
To determine whether a linear ERF accurately describes the association between the pollutants and self-rated health, we estimated a smooth function ERF using a flexible Bayesian p-spline implemented with a second-order random walk latent field model as described in Section 3.1.1. We estimated these potentially non-linear ERFs using the SPAR-Forest-ERF model, because the simulation study showed it produced the most accurate estimates. Non-linear ERFs were estimated for both PM$_{2.5}$ and PM$_{10}$ with both bad and good self-rated health outcomes, while NO$_{2}$ was not considered because most of its linear ERFs showed no significant associations (see Table \ref{table_RR}). 

\begin{figure}
    \centering
    \includegraphics[width=1\linewidth]{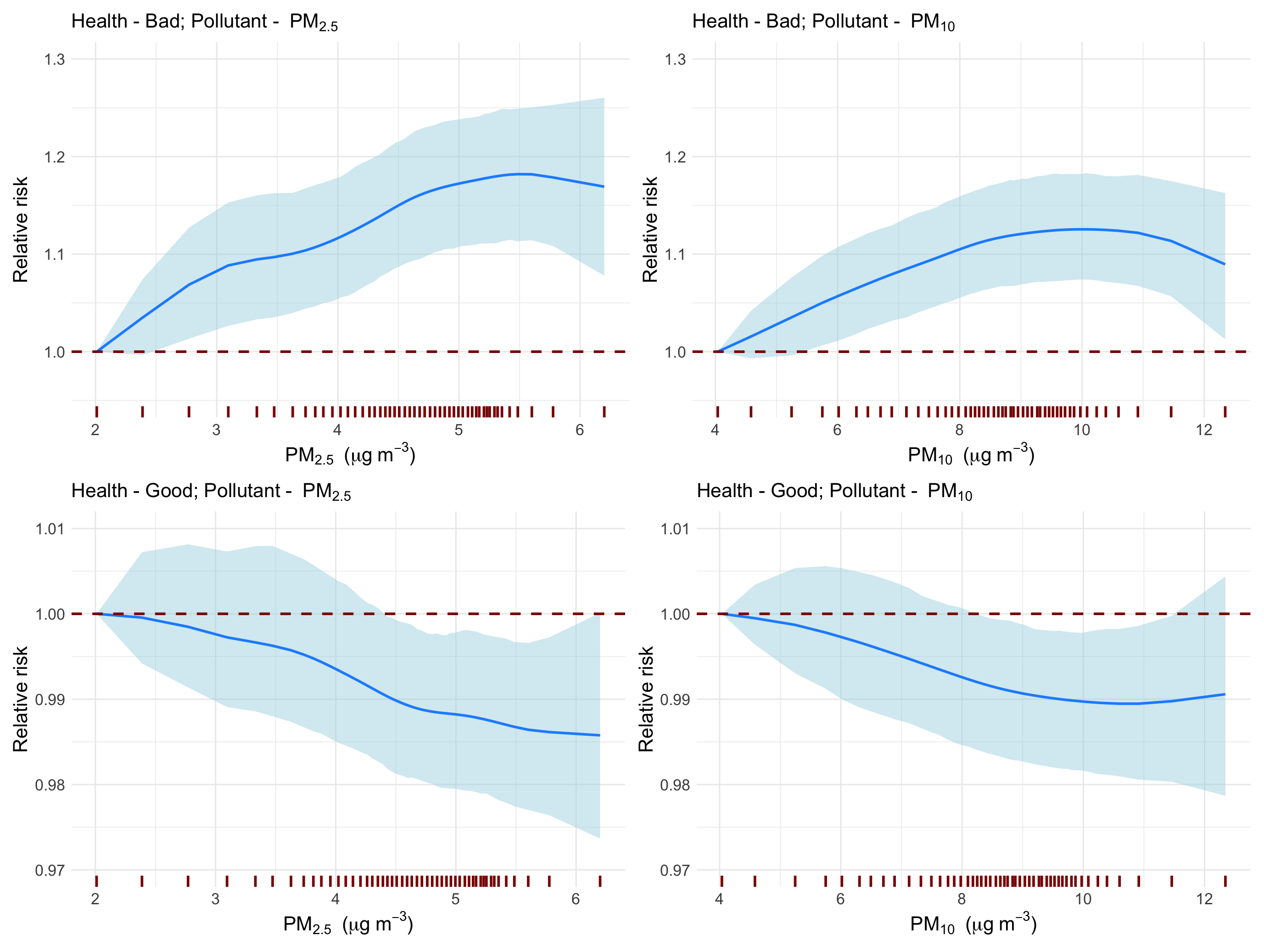}
    \caption{Estimates and pointwise 95\% credible intervals for the exposure response functions $g[x(\mathcal{S}_k)]$ for PM$_{2.5}$ (left) and PM$_{10}$ (right) for bad (top) and good (bottom) self-rated health from the SPAR-Forest-ERF model. The ERFs are presented as RRs relative to the minimum pollutant concentration observed in the study. The red dashed line at a relative risk of one corresponds to no association.}
    \label{fig:nonlinearERF}
\end{figure}

The four estimated ERFs with pointwise 95\% credible intervals are displayed in Figure \ref{fig:nonlinearERF}, where the left and right columns relate to PM$_{2.5}$ and PM$_{10}$ respectively, while the top and bottom rows relate to the bad and good health outcomes. Each ERF is presented as an RR relative to the minimum pollutant concentration observed in the study, which is why the RR for that concentration equals one with no posterior uncertainty. Finally, the red dashed line at a relative risk of one represents no association. The figure shows that the ERFs exhibit generally linear shapes for low to moderate concentrations, followed by a plateau at the highest concentrations observed in the study.  However, a completely linear ERF fits within the pointwise 95\% credible intervals in all four cases, suggesting that there is no conclusive evidence against a linear ERF. The two ERFs for bad health are significantly different from the null relative risk of one for almost all concentrations, showing consistent evidence of an association. In contrast, for the good health outcome, while a constant relative risk of one is not supported by the data, the pointwise 95\% credible intervals contain one at  low to moderate and again at the very highest concentrations, suggesting a much weaker relationship.  

\section{Discussion}
This paper proposes the SPAR-Forest-ERF algorithm for population-level epidemiological risk assessment, which fuses the power of random forests for capturing non-linear and interacting confounder-response effects with Bayesian spatial smoothing models that can estimate interpretable relative risk based exposure response functions with full UQ while capturing the residual spatial autocorrelation ubiquitous in such data. The algorithm is the first fusion of machine learning algorithms and Bayesian spatial smoothing models for areal unit data where the goal of the analysis is effect estimation rather than prediction. Methodologically, the algorithm builds on \cite{macbride2025} and \cite{figueira2025} by correctly propagating uncertainty between both the ML and statistical models, developing a new stopping criteria designed  to ensure the stability of  the primary inferential target, and incorporating a range of different ERFs for maximum model flexibility. The widespread usability of the algorithm is enabled by its accompanying software, while its impact could be substantial given its superior performance compared to existing methods and the widespread use of spatial ecological association studies in fields such as economics, environmental epidemiology and social science to name just 3. 

The simulation study provided a number of key messages about the relative efficacy of SPAR-Forest-ERF compared with a generalized linear mixed model with linear confounder-response associations that is commonly used for epidemiological risk assessment. Firstly, in the realistic situation where the exposure is spatially autocorrelated (e.g., air pollution) and the confounders do not exhibit exactly linear relationships with disease risk with no interactions, then the proposed iterative algorithm is vastly superior, with  between 49\% and 83\% better point estimation as measured by the root mean square error. Uncertainty quantification is also superior,  as measured by the 95\% credible intervals being narrower but having percentage coverages closer to the nominal 95\% level. These findings appear to be consistent across different shaped ERFs (e.g., linear and sigmoidal), which makes SPAR-Forest-ERF an excellent general purpose tool for quantifying the effect of an exposure on a response using spatial areal unit data.  We note that even in the unlikely idealized situations where: (i) all confounders exhibit exactly linear relationships with the response with no interactions; or (ii) the exposure is completely independent in space and of the confounders; then SPAR-Forest-ERF performs similarly to the commonly used generalized liner mixed model, particularly in point estimation. We further highlight that the iterative nature of our algorithm is necessary, because the simulation study results show that a one-iteration simplification of our algorithm performs worse in almost all scenarios. 

The results from the motivating study show that particulate matter air pollution measured as PM$_{2.5}$ and PM$_{10}$ exhibit consistent significant associations with general self-rated health, which agrees with the vast epidemiological literature quantifying its effects on specific morbidity (e.g., \citealp{forastiere2024}) and mortality (e.g., \citealp{orellano2024}) end-points. The robustness of the evidence from this study is strong due to the consistency of the findings from the range of analyses conducted, which include linear and non-linear ERFs, the allowance for exposure measurement error, and the assessment of the impact on both bad and good self-rated health.  The study also adds to the growing body of evidence such as \cite{boogaard2024} suggesting that air pollution is harmful even at low concentrations,  specifically those below the UK National Air Quality Strategy Objectives and the World Health Organization Air Quality Guidelines. For example, PM$_{2.5}$ was found to significantly increase the risk of self-rated bad health by between 2.3\% and 4.1\% for a 1$\mu gm^{-3}$ increase in long-term concentrations, despite all Data Zones in Scotland having concentrations below the UK objective of 10$\mu gm^{-3}$ and 68\% of DZs being below the WHO guideline of 5$\mu gm^{-3}$. 

Additionally, the magnitude of the estimated relative risks associating air pollution with self-rated good health are much smaller in magnitude by a factor of around 10 compared to those for self-rated bad health (a 0.4\% change compared to a 4\% change). One possible cause of these differential associations is that there is much less spatial variation in the risk of good health compared to the risk of bad health, which can be observed in the SMRs in Table 1 in the supplementary material. In essence, as the exposure is the same for both outcomes but the scale of the response varies, then the magnitude of the estimated slope coefficient must therefore vary. 

Finally, SPAR-Forest-ERF can be viewed as a generic algorithm that combines a machine learning component for capturing complex confounder-response associations not of direct interest, with a Bayesian smoothing model that estimates interpretable ERFs with full UQ while capturing any residual autocorrelation in the data. As a result, it will be straightforward to modify to suit the needs of other applications, making it a very flexible tool for carrying out risk assessment from spatial data. For example, the spatial autocorrelation model used here could be replaced with a geostatistical style model for point-level data,  or by a spatio-temporal model if the data are available over multiple time periods. Additionally, different types of ERFs could be incorporated such as varying coefficient models, while the random forest component could be replaced by alternative ML algorithms such as neural networks.

\section{Disclosure statement}\label{disclosure-statement}

No conflicts of interest

\section{Data Availability Statement}\label{data-availability-statement}

The data for the motivating study are all publicly available, and their sources are given in Section 2.

\phantomsection\label{supplementary-material}
\bigskip

\begin{center}

{\large\bf SUPPLEMENTARY MATERIAL}

\end{center}

\begin{description}
\item[Supplementary material:]
Supplementary materials available on request. Additional data analyses and methodological description not included in the main paper due to space constraints. (.pdf)
\end{description}

\bibliography{Lee.bib}

\end{document}